\documentclass[12pt]{article}
\pdfoutput=1
\usepackage{epsfig}
\usepackage{amsfonts}
\usepackage{amssymb}
\usepackage{xcolor}

\begin{document}
\newcommand{\nc}{\newcommand}
\nc{\beq}{\begin{equation}} \nc{\eeq}{\end{equation}}
\nc{\beqa}{\begin{eqnarray}} \nc{\eeqa}{\end{eqnarray}}
\nc{\eps}{{\epsilon}}
\nc{\R}{{\cal R}}
\nc{\A}{{\cal A}}
\nc{\K}{{\cal K}}
\nc{\B}{{\cal B}}
\nc{\C}{{\cal C}}
\begin{center}

{\bf \Large  The Structure  of UV Divergences in Maximally\\[0.4cm] Supersymmetric Gauge Theories } \vspace{1.0cm}

{\bf \large D. I. Kazakov$^{1,2}$, A. T. Borlakov$^{1,2}$, D. M. Tolkachev$^{1,2,3}$\\[0.3cm] and D. E. Vlasenko$^{4}$ }\vspace{0.5cm}

{\it
$^1$Bogoliubov Laboratory of Theoretical Physics, Joint
Institute for Nuclear Research, Dubna, Russia.\\
$^2$Moscow Institute of Physics and Technology, Dolgoprudny, Russia\\
$^3$Stepanov Institute of Physics, Minsk, Belarus\\
$^4$Department of Physics, South Federal State University, Rostov-Don, Russia}
\vspace{0.5cm}

\abstract{We consider the UV divergences up to sub-sub leading order for the four-point on-shell scattering amplitudes in D=8 supersymmetric Yang-Mills theory in the planar limit.  We trace how the leading, subleading, etc divergences appear in all orders of perturbation theory.  The structure of these divergences is typical for any local quantum field theory independently on renormalizability. We show how the generalized RG equations  allow one to evaluate the leading, subleading, etc. contributions in all orders of PT starting from one-, two-, etc. loop diagrams respectively. We focus  then on subtraction scheme dependence of the results and show that in full analogy with renormalizable theories the scheme dependence can be absorbed into the redefinition of the couplings. The only difference is that the role of the couplings play dimensionless combinations like $g^2s^2$ or $g^2t^2$, where $s$ and $t$ are the Mandelstam variables.}
\end{center}

Keywords: Amplitudes, maximal supersymmetry, UV divergences

\section{Introduction}

In recent years maximally supersymmetric gauge theories attracted much attention and served as a theoretical playground promising new insight in to the nature of gauge theories beyond usual perturbation theory. 
This became possible  due to the development of new computational techniques such as the spinor helicity  and the on-shell momentum superspace formalism~\cite{Reviews_methods}. The most successful examples are the $\mathcal{N}=4$ SYM theory   in $D=4$ ~\cite{BDS4point3loop_et_all}   and the $\mathcal{N}=8$ SUGRA~\cite{N=8SUGRA finiteness}.  These theories are believed to possess several remarkable properties, among which are total or partial cancelation  of UV divergences, factorization of higher loop corrections and possible integrability. The success of factorization leading to the BDS
ansatz~\cite{BDS4point3loop_et_all} for the amplitudes in $D=4$ $\mathcal{N}=4$ SYM stimulated similar activity in other models and dimensions~\cite{Reviews_Ampl_General}. 
The universality of the developed methods allows one  to apply them to  SYM theories in dimensions higher than 4 \cite{GeneralDimentions,SpinorHelisity_extraDimentions}.  

In this paper,  we focus on the on-shell 4-point amplitude as the simplest structure and  analyze the UV divergences in maximally  SYM theories in $D=8$ dimensions in all loops. This theory has no IR divergences even on shell but
since the gauge coupling $g^2$ here has dimension $-4$,  it is non-renormalizable by power counting. 

\begin{figure}[htb]
 \begin{tabular}{c}
 $\frac{\mathcal{A}_4}{\mathcal{A}_4^{(0)}}=1+\sum\limits_L M^{(L)}_4(s,t)=$  \\
\includegraphics[scale=0.35]{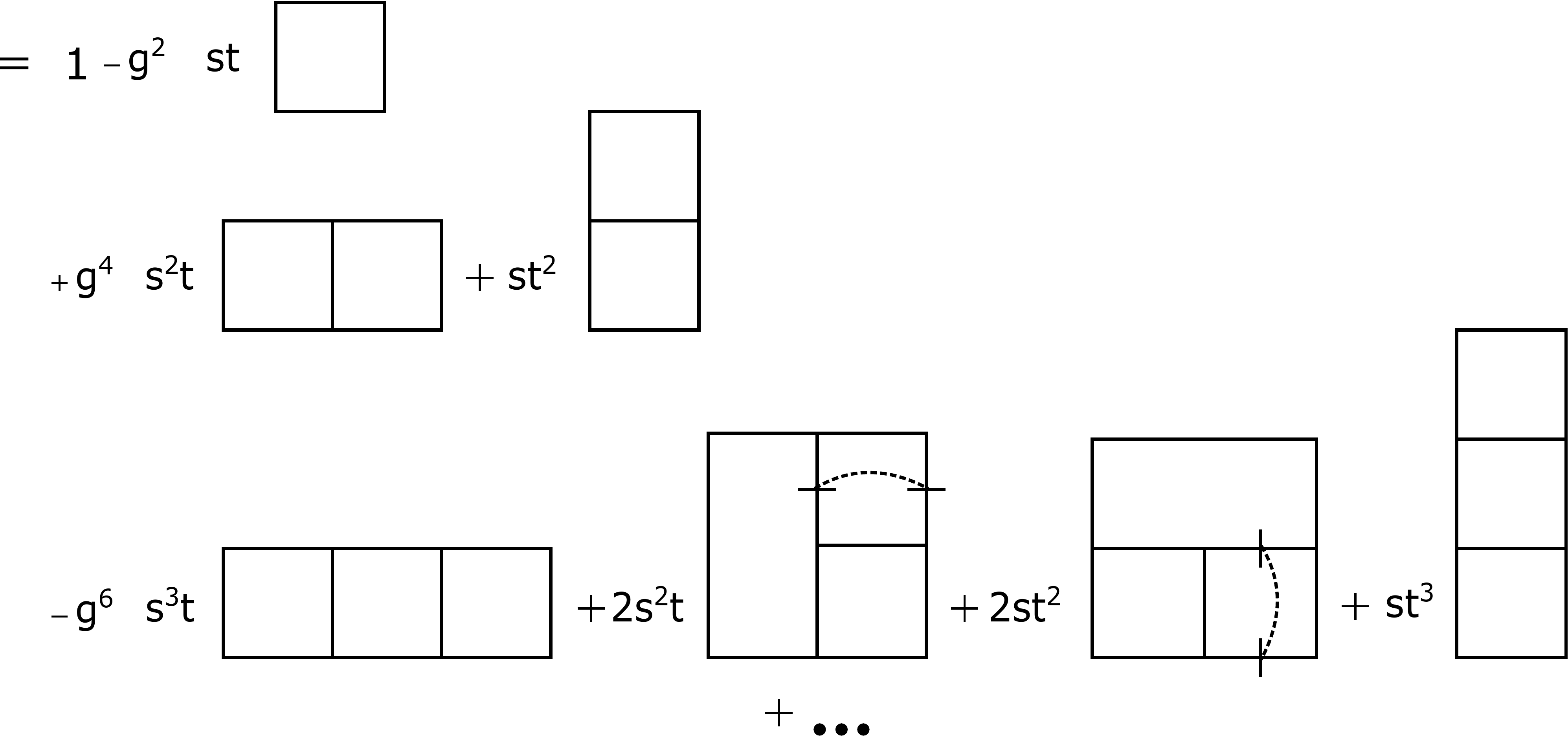}
\end{tabular}
 \caption{The universal expansion for the four-point scattering amplitude in SYM theories in terms of master integrals.
 The connected strokes on the lines mean  the square of the flowing momentum.}\label{expan}
 \end{figure}

Applying first the color decomposition of the amplitudes, we are left with the partial amplitudes. Within the spinor-helicity formalism the tree level partial amplitudes  depend on the Mandelstam variables $s, t$ and $u$ and have a relatively simple universal form. The advantage of the superspace formalism is that the tree level amplitudes always factorize  so that the ratio of the loop corrections to the tree level expression can be expressed in terms of pure scalar master integrals shown in Fig.\ref{expan}~\cite{Bern:2005iz}.

 Within the dimensional regularization (dimensional reduction) the UV divergences manifest themselves as the pole terms with the numerators being the polynomials over the kinematic variables. 
In D-dimensions the first UV divergences start from L=6/(D-4)  loops, consequently, in  D=8 and they start already at one loop.  
Notice that all simple loops as well as triangles completely cancel in all orders of PT. This is the consequence  of maximal supersymmetry and it seems this is maximal it can do. In D=4 this leads to the the cancellation of all the UV divergences since boxes are finite, however, in higher dimensions the UV divergences remain non-renormalizable.

In  recent papers~\cite{we1,we2,we3}, we considered the leading and subleading UV divergences of the on-shell scattering amplitudes for all three cases of maximally supersymmetric SYM theories, D=6 (N=2 SUSY), D=8 (N=1 SUSY) and D=10 (N=1 SUSY). We obtained the recursive relations  that allow one to get the leading and subleading divergences in all loops in a pure algebraic way. Then we constructed the differential equations which are the generalization of the RG equations for non-renormalizable theories. Similar to the renormalizable theories, these equations lead to summation of the leading (and subleading) divergences in all loops. In paper \cite{we4} we concentrated on solving these equations. 


In this paper we summarize all previous results with addition of the sub-subleading case and focus on  the scheme dependence of the counter terms.  We consider the transition from the minimal to non-minimal subtraction scheme and show that it is equivalent to the redefinition of the couplings played by dimensionless combinations $g^2s^2$ or $g^2t^2$. This redefinition, however, differs from a simple multiplication due to the dependence on kinematic factors. When integrated inside the diagrams this factors lead to the more complicated procedure which manifests itself already in the recurrence relations.

\section{Recurrence relations for the leading, subleading and sub-subleading divergences in D=8 N=1 SYM theory }

Any local quantum field theory has a remarkable property that after performing the
incomplete ${\cal R}$-operation, the so-called ${\cal R}^\prime$-operation, the remaining UV divergences are always local. This property allows one to construct the so called recurrence relations which relate the divergent contributions in all orders of perturbation theory (PT) with the lower order ones.  In renormalizable theories this relations are known as pole equations (within dimensional regularization) and are governed by the renormalization group \cite{hooft}. The same is true though technically is more complicated in any local theory as we have demonstrated in \cite{we2, we3}.  We remind here some features of this procedure.

The incomplete ${\cal R}$-operation (${\cal R}^\prime$-operation) subtracts only the subdivergences of a given graph, while the full R operation is defined as 
\begin{equation}
\R G = (1-\K) \R' G,
\end{equation}
where ${\cal K}$ is an operator that singles out the singular part of the graph and $K{\cal R}^\prime G$- is the counter term corresponding to the graph G.
After applying the ${\cal R}^\prime$-operation to a given graph in the n-th order of PT one gets the following series of therms
\beqa
{\cal R'}G_n&=&\frac{\A_n^{(n)}(\mu^2)^{n\epsilon}}{\epsilon^n}+\frac{\A_{n-1}^{(n)}(\mu^2)^{(n-1)\epsilon}}{\epsilon^n}+ ... +\frac{\A_1^{(n)}(\mu^2)^{\epsilon}}{\epsilon^n}\nonumber \\
&+&\frac{\B_n^{(n)}(\mu^2)^{n\epsilon}}{\epsilon^{n-1}}+\frac{\B_{n-1}^{(n)}(\mu^2)^{(n-1)\epsilon}}{\epsilon^{n-1}}+ ... +\frac{\B_1^{(n)}(\mu^2)^{\epsilon}}{\epsilon^{n-1}} \nonumber \\
&+&\frac{\C_n^{(n)}(\mu^2)^{n\epsilon}}{\epsilon^{n-2}}+\frac{\C_{n-1}^{(n)}(\mu^2)^{(n-1)\epsilon}}{\epsilon^{n-2}}+ ... +\frac{\C_1^{(n)}(\mu^2)^{\epsilon}}{\epsilon^{n-2}} \nonumber\\
&+&\mbox{lower\ pole\ terms,}\label{Rn}
\eeqa
where the terms like $\frac{\A_{k}^{(n)}(\mu^2)^{k\epsilon}}{\epsilon^n}$  or $\frac{\B_{k}^{(n)}(\mu^2)^{k\epsilon}}{\epsilon^{n-1}}$ come from the $k$-loop graph which survives after subtraction of the $(n-k)$-loop counterterm.
The resulting expression has to be local, hence do not contain terms like $\log^l{\mu^2}/\epsilon^k$, from any $l$ and $k$. This requirement leads to the sequence of relations for $\A_i^{(n)}, \B_i^{(n)}$ and $\C_i^{(n)}$ which can be solved in favour of the lowest order terms
\beqa
\A_n^{(n)}&=&(-1)^{n+1}\frac{\A_1^{(n)}}{n}, \nonumber\\
\B_n^{(n)}&=&(-1)^n \left(\frac 2n \B_2^{(n)}+\frac{n-2}{n}\B_1^{(n)}\right) \label{rel},\\
C_n^{(n)}&=&(-1)^{n+1}\left(\frac{3}{n}C_3^{(n)}+\frac{2(n-3)}{n}C_2^{(n)}+\frac{(n-2)(n-3)}{2n}C_1^{(n)}\right).\nonumber \label{abc_coeff}
\eeqa
It is useful also to write down the local expression for the ${\cal KR'}$ terms (counter terms) equal to
\beq
{\cal KR'}G_n=\sum_{k=1}^n \left(\frac{\A_k^{(n)}}{\epsilon^n} +\frac{\B_k^{(n)}}{\epsilon^{n-1}}+\frac{\C_k^{(n)}}{\epsilon^{n-2}}\right)\equiv
\frac{\A_n^{(n)'}}{\epsilon^n}+\frac{\B_n^{(n)'}}{\epsilon^{n-1}}+\frac{\C_n^{(n)'}}{\epsilon^{n-2}}.
\eeq
Then one has, respectively
\beqa
\A_n^{(n)'}&=&(-1)^{n+1}\A_n^{(n)}=\frac{\A_1^{(n)}}{n}, \nonumber \\
\B_n^{(n)'}&=& \left(\frac{2}{n(n-1)} \B_2^{(n)}+\frac{2}{n}\B_1^{(n)}\right) \label{rel2},\\
C_n^{(n)'}&=&\left(\frac{2}{(n-1)(n-2)}\frac{3}{n}C_3^{(n)}+\frac{2}{n-1}\frac{3}{n}C_2^{(n)}+\frac{3}{n}C_1^{(n)}\right). 
\nonumber
\eeqa
	
This means that performing the  ${\cal R}'$-operation one can take care only of the one-, two-, three-loop  diagrams surviving after contraction and get the desired leading pole terms via eq.(\ref{rel})  in the leading, subleading and sub-subleading order, respectively. They can be calculated  in all loops pure algebraically. 

\begin{figure}[htb!]
\begin{center}
\leavevmode
\includegraphics[width=0.7\textwidth]{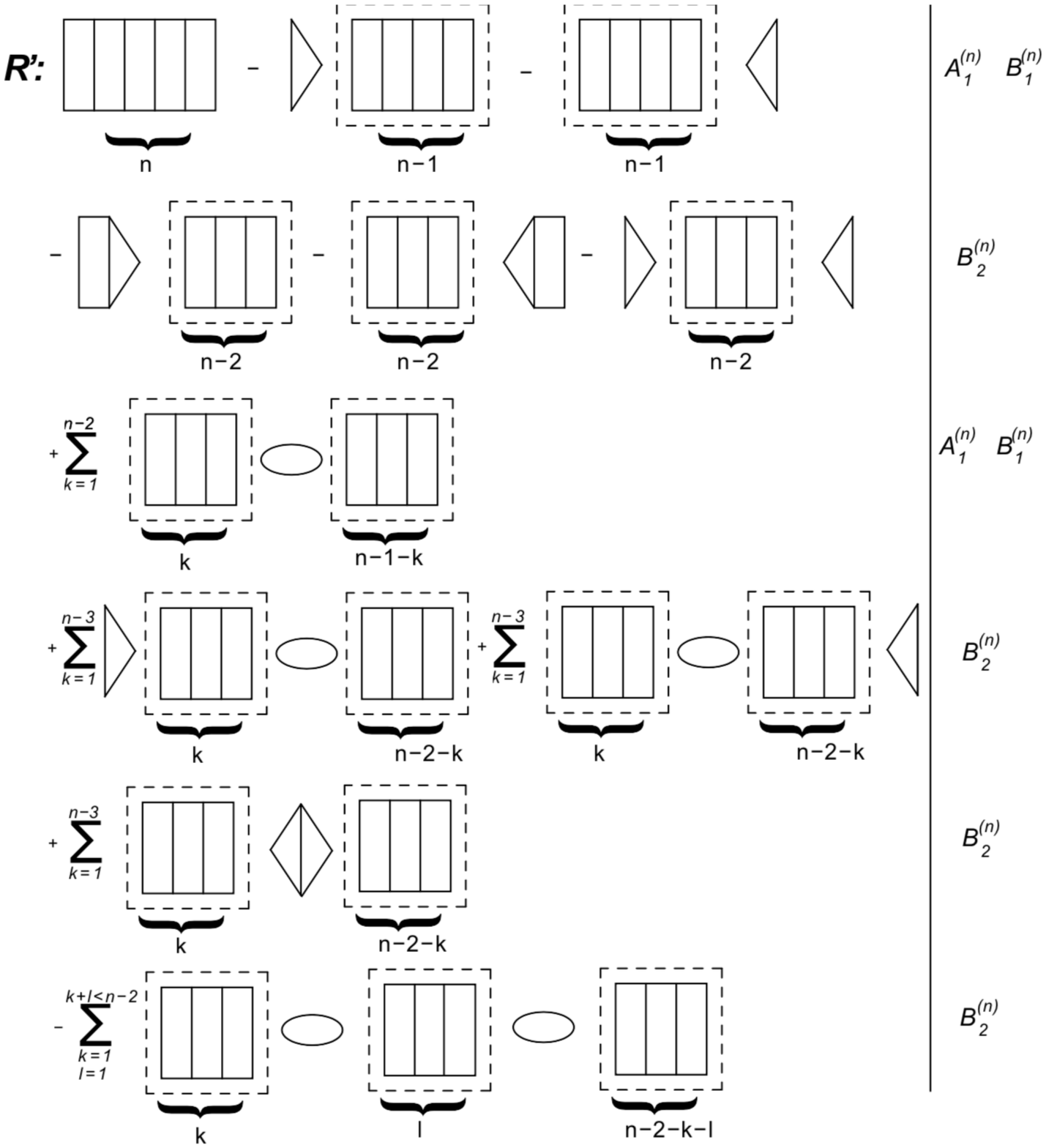}
\end{center}\vspace{-1cm}
\caption{$\R'$-operation for the horizontal ladder in D=8}
\label{R}
\end{figure}

Remind how this procedure works in case of the ladder-type diagrams  shown in Fig.\ref{R} \cite{we3}. 
Consider first the leading order. Since the horizontal ladder-type diagrams in the leading order depend only on $s$ further on we simplify the notation $ A_n^{(n)}=s^{n-1}A_n$ and $ A_n^{(n)'}=s^{n-1}A'_n$. Calculating now the one-loop diagrams shown in the first and third raws of Fig.\ref{R} and substituting them into eq.(\ref{rel}) we obtain the recurrence relation in the leading order
%
\beq
n A_n=-\frac{2}{4!} A_{n-1}+\frac{2}{5!}\sum_{k=1}^{n-2}A_kA_{n-1-k}, \ \ \  n\geq 2, \label{one8}
\eeq
where $A_1=1/3!$. Using this recurrence relation one can calculate the leading divergence in any loop order starting from the one-loop one pure algebraically.

In subleading order one has the terms linear in $t$. To separate them  we use the notation
 $B_n^{(n)}=s^{n-1}B_{ns-1}+s^{n-2}tB_{tn}$ and $B_n^{(n)'}=s^{n-1}B^{'}_{ns-1}+s^{n-2}tB^{'}_{tn}$. 
 To get the recurrence relation in subleading case one has to calculate the two-loop diagrams shown in the second and the last raws of Fig.\ref{R}. We start with the primed quantities since they actually enter the recurrence relations
 \beqa
B'_{tn}&=&-\frac{2}{n(n-1)}B'_{tn-2}\frac{10}{5!5!}+\frac 2n B'_{tn-1}\frac{2}{5!}, \label{btprime} \\
B'_{sn}&=&\frac{2}{n(n-1)}\left[-A'_{n-2}\frac{2321}{5!5!2}-B'_{sn-2}\frac{18}{4!5!}+B'_{tn-2}\frac{44}{5!5!}\right. \nonumber\\
&-& \left. \sum_{k=1}^{n-3}A'_kA'_{n-2-k}\frac{938}{4!5!15}- \sum_{k=1}^{n-3}A'_kB'_{sn-2-k}\frac{1}{5!2}+
 \sum_{k=1}^{n-3}A'_kB'_{tn-2-k}\frac{442}{5!5!12}\right. \nonumber\\
 &-&\left. \sum_{k,l=1}^{n-k+l<n-2}A'_kA'_lA'_{n-2-k-l}\frac{8}{5!5!}\frac{46}{15}
  -\sum_{k,l=1}^{n-k+l<n-2}A'_kA'_lB'_{sn-2-k-l}\frac{12}{5!5!}\right. \nonumber \\
 &+&\left. \sum_{k,l=1}^{n-k+l<n-2}A'_kA'_lB'_{tn-2-k-l}\frac{4}{5!5!}
  +\sum_{k,l=1}^{n-k+l<n-2}B'_kA'_lA'_{sn-2-k-l}\frac{2}{5!5!}\right]\nonumber \\
  &+&\frac 2n\left[ A'_{n-1}\frac{19}{3 4!}+B'_{sn-1}\frac{2}{4!}-B'_{tn-1}\frac{4}{5!}\right. \nonumber \\
 &+& \left. \sum_{k=1}^{n-2}A'_kA'_{n-1-k}\frac{2}{5!}\frac{46}{15}+\sum_{k=1}^{n-2}A'_kB'_{sn-1-k}\frac{4}{5!}-
 \sum_{k=1}^{n-2}A'_kB'_{tn-1-k}\frac{2}{5!}\right] .\label{bsprime}
\eeqa
where $B'_{s1}=B'_{t1}=0$, $B'_{s2}=-5/3!/4!/12$, $B'_{t2}=-1/3!/4!/6$. And similar for the unprimed ones. Recurrence relations for the sub-subleading divergences are too lengthy to present them here.

 
Solution of the recurrence relations (\ref{one8},\ref{btprime},\ref{bsprime}) are complicated. However, since we actually need the sum of the series we perform the summation
multiplying both sides of eq.(10) by $z^{n-1}$ and take the sum from 3 to infinity. After some algebraic manipulation, introducing the notation  $\Sigma_A=\sum_{n=1}^\infty A_n (-z)^n$  we finally transform the recurrence relations to differential equations. In the leading order one has (hereafter $z \equiv g^2 s^2/\epsilon$)
\beq
\frac{d}{dz}\Sigma_A=-\frac{1}{3!}+\frac{2}{4!}\Sigma_A-\frac{2}{5!}\Sigma_A^2. \label{eqa}
\eeq

A similar differential  equation can be obtained for $\Sigma'_{sB}=\sum_2^\infty z^nB'_{sn}$ and $\Sigma'_{tB}=\sum_2^\infty z^nB'_{tn}$,
\beq
\frac{d^2 \Sigma'_{tB}(z)}{dz^2}-\frac{1}{30}\frac{d \Sigma'_{tB}(z)}{dz}+\frac{\Sigma'_{tB}(z)}{720}=-\frac{1}{432},
\label{eq1}
\eeq
\beq
\frac{d^2 \Sigma'_{sB}(z)}{dz^2}+f_1(z)\frac{d \Sigma'_{sB}(z)}{dz}+f_2(z)\Sigma'_{sB}(z)=f_3(z), \label{Ric}
\eeq
with
\beqa
f_1(z)&=&-\frac 16+\frac{\Sigma_A}{15},\nonumber\\
f_2(z)&=&\frac{1}{80}-\frac{\Sigma_A}{120}+\frac{\Sigma_A^2}{600}+\frac{1}{15}\frac{d \Sigma_A}{dz}, \nonumber\\
f_3(z)&=&\frac{2321}{5!5!2}\Sigma_A+\frac{11}{1800}\Sigma'_{tB}-\frac{469}{5!90}\Sigma_A^2-\frac{442}{5!5!6}\Sigma_A\Sigma'_{tB}+\frac{23}{6750}\Sigma_A^3+\frac{1}{1200}\Sigma_A^2\Sigma'_{tB}\nonumber\\
&-&\frac{19}{36}\frac{d \Sigma_A}{dz}-\frac{1}{15}\frac{d \Sigma'_{tB}}{dz}
+\frac{23}{225}\frac{d \Sigma_A^2}{dz}+\frac{1}{30}\frac{d( \Sigma_A\Sigma'_{tB})}{dz}-\frac{3}{32}.\nonumber
\eeqa

Solutions to these equations are simple only for the leading order (\ref{eqa}). Indeed solution to eq.(\ref{eqa}) is \cite{we2}
\beq
\Sigma_A(z)=-\sqrt{5/3}\frac{4\tan(z/(8\sqrt{15}))}{1-\tan(z/(8\sqrt{15}))\sqrt{5/3}}.\label{solA}
\eeq
It has infinite number of poles and no limit when $z\to\infty$. In subleading order there is no simple analytic solution,
however, qualitatively it behaves similar to (\ref{solA}) \cite{we3}.

One can construct also similar recurrence relations in general case including all diagrams of PT. 
In \cite{we2}, we constructed the full recurrence relation for the leading divergences. It has been done by consistent application of the $\R'$-operation and integration over the remaining triangle and bubble diagrams with the help of Feynman parameters. Denoting by  $S_n(s,t)$ and  $T_n(s,t)$  the sum of all contributions  in the  $n$-th order of PT in $s$ and  $t$ channels, respectively, we got the following recursive relations:
\beqa
&&nS_n(s,t)=-2 s^2 \int_0^1 dx \int_0^x dy\  y(1-x) \ (S_{n-1}(s,t')+T_{n-1}(s,t'))|_{t'=tx+uy}\label{req8}\nonumber \\ &+&
s^4 \int_0^1\! dx \ x^2(1-x)^2 \sum_{k=1}^{n-2}  \sum_{p=0}^{2k-2} \frac{1}{p!(p+2)!} \
 \frac{d^p}{dt'^p}(S_{k}(s,t')+T_{k}(s,t')) \times \nonumber \\
&&\hspace{2cm}\times  \frac{d^p}{dt'^p}(S_{n-1-k}(s,t')+T_{n-1-k}(s,t'))|_{t'=-sx} \ (tsx(1-x))^p, \label{req8}
\eeqa
where  $S_1= \frac{1}{12},\ T_1=\frac{1}{12}$, $u=-s-t$. 

As in the ladder case,   this recurrence relation takes into account all the diagrams of a given order of PT and allows one to sum all orders of PT. This can be achieved by multiplying both sides of eq.(\ref{req8}) by $(-z)^{n-1}$, where $z=\frac{g^2}{\epsilon}$ and summing up from n=2 to infinity. Denoting the sum by $\Sigma(s,t,z)=\sum_{n=1}^\infty S_n(s,t) (-z)^n$, we finally get the following differential equation
\beqa
&&\frac{d}{dz}\Sigma(s,t,z)=-\frac{1}{12}+2 s^2 \int_0^1 dx \int_0^x dy\  y(1-x)\ (\Sigma(s,t',z)+\Sigma(t',s,z))|_{t'=tx+uy}
\label{eq8}\\
&&-s^4  \int_0^1\! dx \ x^2(1-x)^2 \sum_{p=0}^\infty \frac{1}{p!(p+2)!} (\frac{d^p}{dt'^p}(\Sigma(s,t',z)+\Sigma(t',s,z))|_{t'=-sx})^2 \ (tsx(1-x))^p. \nonumber
\eeqa
The same equations with the replacement $s \leftrightarrow t$ are valid for $\Sigma(t,s,z)$.

As one can see, the equation (\ref{eq8}) is integro-differential and  cannot be treated analytically. Instead, we performed a numerical study of this equation \cite{we4}. The result is shown in Fig.\ref{allloop8}.
\begin{figure}[htb!]
\begin{center}
\leavevmode
\includegraphics[width=0.5\textwidth]{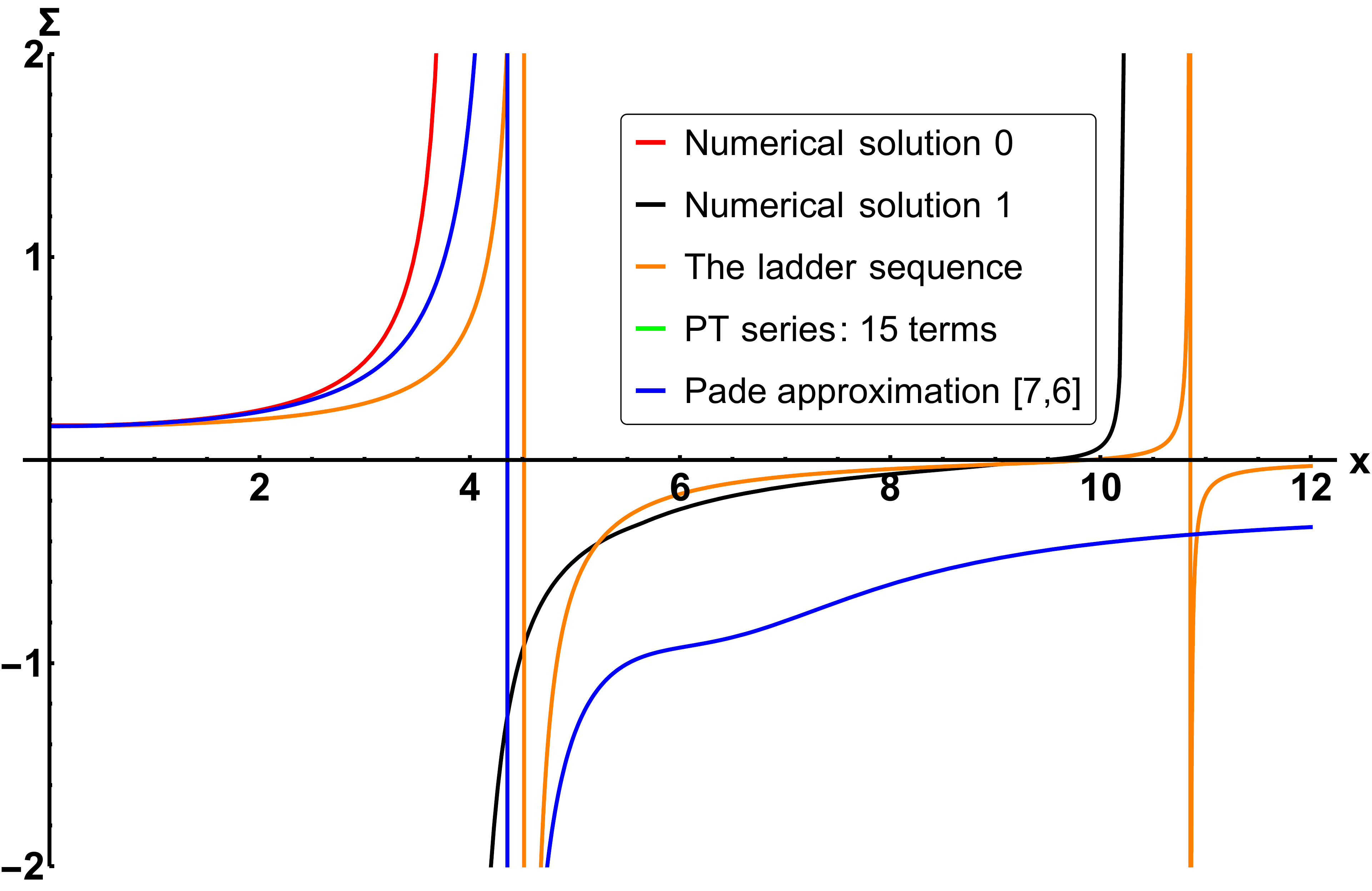}
\label{allloop8}
\end{center}
\caption{Comparison of various approaches to solve eq.(\ref{eq8}) . The red and black lines are the numerical solutions described in the previous section between before the first pole and between the first and the second ones. The green one is the PT. The blue one is the Pade approximation. And last one is yellow which represents the Ladder analytical solution.}
\label{allloop8}
\end{figure}
One can see that all the curves practically have the same behaviour. It is clearly seen that the numerical curve reproduces both poles and is close to the ladder approximation. This comparison justifies our conclusion that the ladder approximation reproduces the correct behaviour of the function. 

Our analysis shows that in the leading and subleading orders summation of the UV divergences leads to the sum which is a function with infinite number of poles for any choice of kinematics. This function has no limit when $z\to\infty$ ($\epsilon\to 0$). This means that  the UV finiteness is not reached when the sum over all loops is taken into account. This limit would correspond to removing the UV regularization. One can see that summation of the whole infinite series does not improve the situation. One can not just remove the UV regularization and get a finite theory.

\section{The scheme dependence}

The problem with non-renormalizable interactions is not that the scattering amplitudes can not be made finite. After all one can subtract all UV divergences in a minimal way. The problem is that  the structure of the counter terms does not repeat the original Lagrangian and one gets new structures with increasing power of momenta at each step of perturbation theory. 
This means that subtracting the UV divergence each time, one has to define the normalization of a new operator, thus having a new arbitrary constant.  The number of these constants is infinite.  However, as we have found out, all the higher order divergences are related via the generalized RG equations. This means that the above mentioned arbitrariness of the counter terms, and hence of the finite parts, is also restricted and one may hope to relate them. In what follows we study this problem and consider the arbitrariness in the counter terms that appears when going from the minimal to non-minimal subtraction scheme.

\subsection{The subleading case}

All the calculations presented so far were based on the minimal subtraction scheme. Obviously, the leading divergences are scheme independent but the subleading ones depend on a scheme.  However, this dependence in all orders of PT is defined by a single arbitrary constant. Indeed, all the recurrence relations obtained above are scheme independent!  The only dependence on the subtraction scheme  is contained in subtraction of a single one-loop box-type diagram. If one chooses the one-loop counter term in the form 
\beq
A_1'+B_{s1}'=\frac{1}{6\epsilon}(1+c_1 \epsilon)
 \eeq
 ($c_1=0$ corresponds to the minimal subtraction scheme), then using the recurrence relations for the subleading divergences, one gets the following additional term to the sum of the counter terms in all orders of PT (remind the notation $z\equiv g^2s^2/\epsilon$)
  \beq	
 \Delta \Sigma_{sB}'=c_1 z\frac{d\Sigma'_A}{dz}.
 \label{add}
 \eeq
Thus, the arbitrariness in the definition of the counter terms with an infinite number of derivatives is reduced in the leading order to the choice of the single parameter $c_1$. 
It is equivalent to renormalization of the coupling constant in the following form:
 \beq	
 z\rightarrow z (1+c_1\epsilon).
 \label{add}
 \eeq
This is exactly what happens with renormalizable interactions except that the coupling $g^2$ here has dimension $-4$
and one has to choose the  dimensionless combination $g^2s^2$. Obviously, keeping both $s$ and $t$, we also have  $g^2t^2$. Hence, one can not just say that the change of the subtraction scheme is equivalent to the redefinition of a single coupling $g^2$, instead there are two of them and this redefinition depends on kinematics.


\subsection{The sub-subleading case}

Consider now what happens in the  sub-subleading order. 
In this case, the dependence on the subtraction scheme is contained also in the two-loop box-type diagram. Following the subleading case, we choose the counter term in the form 

\begin{equation}
A_2'+B_2'=\frac{s}{3!4!\epsilon^2}\left(1-\frac{5}{12}\epsilon+2c_1\epsilon+c_2\epsilon^2\right),
\end{equation}
where $c_1$ comes from the one-loop counter term and $c_2$ is the new subtraction constant. Using the recurrence relation for the sub-subleading divergences, one gets the following additional term  proportional to $c_2$ in all orders
of PT
\beq	
 \Delta \Sigma_{sC}'=c_2 z^2\frac{d\Sigma_A'}{dz}.
 \eeq
This corresponds to the shift of the coupling constant (we put here $t=0$ for simplicity)
 \beq	
 z\rightarrow z(1+c_1\epsilon) + z^2c_2 \epsilon^2.  
 \label{add2}
 \eeq
This simple pattern obviously has a one-loop origin since it comes from the leading divergences and they are defined by 
the one-loop box diagram.

The situation with dependence on $c_1$ in the sub-subleading order  is more complicated. There are two contributions here: the linear and quadratic. The quadratic dependence obviously appears from the substitution of expression (\ref{add}) into the
minimal scheme counter term $\Sigma'_A$, which gives the second derivative of $\Sigma'_A$.  However, this is not the only contribution. The redefinition of the coupling in fact contains  an extra part  compared to (\ref{add2}) which is proportional to  $c_1^2$
 \beq	
 z\rightarrow z(1+c_1\epsilon) + z^2(c_2 +c_1^2/4!)\epsilon^2.  
 \label{add3}
 \eeq
It gives the first derivative of $\Sigma'_A$. All together the full quadratic dependence has the form
\beq 
 \Delta \Sigma_{sC}'=-c_1^2\frac{z}{4!}\left(\frac{d\Sigma'_A}{dz}-12 \frac{d^{2}\Sigma'_A}{dz^{2}}\right).\label{two}
\eeq
Using the recurrence relations in the sub-subleading order, we have checked that this result is valid in all orders of PT.

This dependence on $c_1^2$ seems to have a general nature valid also
for the $c_1^n$ contributions in the next orders.
To check it, we  calculated the first terms proportional to $c_1^3$ using the  $\R'$-operation in 5-, 6- and 7-loop ladder type box diagrams. The result is
\beq 
\Delta R'_{5boxes}=\frac{s^4}{777600 \epsilon^2}c_1^3,
\eeq
\beq 
\Delta R'_{6boxes}=\frac{s^5}{4665600 \epsilon^3}c_1^3,
\eeq
\beq 
\Delta R'_{7boxes}=\frac{11s^6}{447897600 \epsilon^4}c_1^3.
\eeq
And though we do not have the all loop recurrence relation in this case, the equations written above suggest the following general expression:
\beq 
 \Delta \Sigma_{sC}'=c_1^3\frac{z}{6!} \left(\frac{d\Sigma'_A}{dz}-30 \frac{d^{2}\Sigma'_A}{dz^{2}}+120 \frac{d^{3}\Sigma'_A}{dz^{3}} \right).\label{three}
\eeq
We checked also the $c_1^4$ term
\beq 
\Delta R'_{7boxes}=\frac{s^6}{13996800 \epsilon^3}c_1^4,
\eeq
\beq 
\Delta R'_{8boxes}=\frac{s^7}{671846400 \epsilon^4}c_1^4,
\eeq
and conject a similar expression
\beq 
 \Delta \Sigma_{sC}'=-c_1^4\frac{z}{4!6!} \left(\frac{d\Sigma'_A}{dz}-78 \frac{d^{2}\Sigma'_A}{dz^{2}}+720 \frac{d^{3}\Sigma'_A}{dz^{3}}-1440  \frac{d^{4}\Sigma'_A}{dz^{4}}\right).\label{three}
\eeq

The situation with the linear term is not that straightforward. It is not given by the leading term only but involves also the subleading one. And since the subleading terms depend not only on $s$ but also on $t$, one cannot ignore the $t$-dependence anymore.
It’s clearly seen in the third order of PT. Namely, if we consider the $\R'$-operation of the 3-loop box diagram
\begin{figure}[htb]
\begin{center}
\includegraphics[scale=0.8]{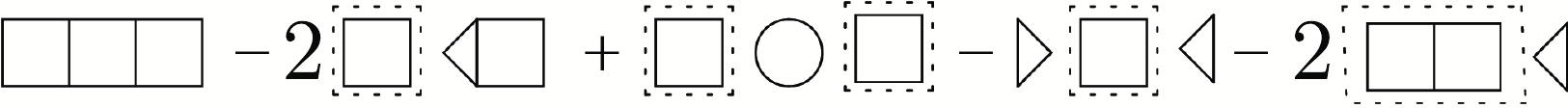}
 \caption{$\R'$-operation for the 3-loop box diagram}\label{3_loop}
 \end{center}
 \end{figure}
in the minimal and non-minimal schemes and calculate the arbitrariness $\Delta \Sigma_{sC}'$, the latter is independent of the $t$ contribution. The reason is that while the two loop box contains the $t$ contribution in the subleading order, the arbitrariness is contained only in the $s$ term. At the same time, when one evaluates the sub-subleading divergence in the 3-loop box diagram using the $\R'$-operation, one has a nonzero contribution from both the $s$ and $t$ terms in the last diagram in Fig.\ref{3_loop}. The two expressions are obviously different
\beq	
\Delta \Sigma_{sC}'(3-loop) = - \frac{719 c_1 s^2}{1036800 \epsilon}, 
\label{delta_3}
\eeq
 whereas $\Sigma_{sB}'$ in 3 loops has the following form:
\beq 
\Sigma_{sB}'(3-loop) = - \frac{71 s^2}{345600\epsilon^2}.\label{sB_3}
\eeq

To take care of this missing $t$ contribution, one has to consider the general case when both the $s$ and $t$ dependences of the amplitude are left.
We, however, suggest proceeding in a different way.  We subtract the unmatched $t$ contribution from $\Sigma_{sB}'$ and compare it with $\Delta \Sigma_{sC}'$. We call it $\Sigma_{sB}^{'trunc}$.
Introducing the initial data $bs'[2] = (2*5sc_1)/1728$ into the recurrence relation at 2 loops 
and excluding the contribution of the $t$ term from $\Sigma_{sB}'$, we found out that in the third order
\beq 
 \Sigma_{sB}^{'trunc}(3-loop) = - \frac{719 s^2}{3110400 \epsilon^2}. \label{t_prime}
\eeq
Taking the derivative with respect to $z$, one reproduces the desired result
\beq	
 \Delta \Sigma_{sC}'(3-loop)=c_1 z\frac{d\Sigma_{sB}^{'trunc}}{dz}(3-loop).
 \eeq

\begin{figure}[htb]
\begin{center}
\includegraphics[scale=0.8]{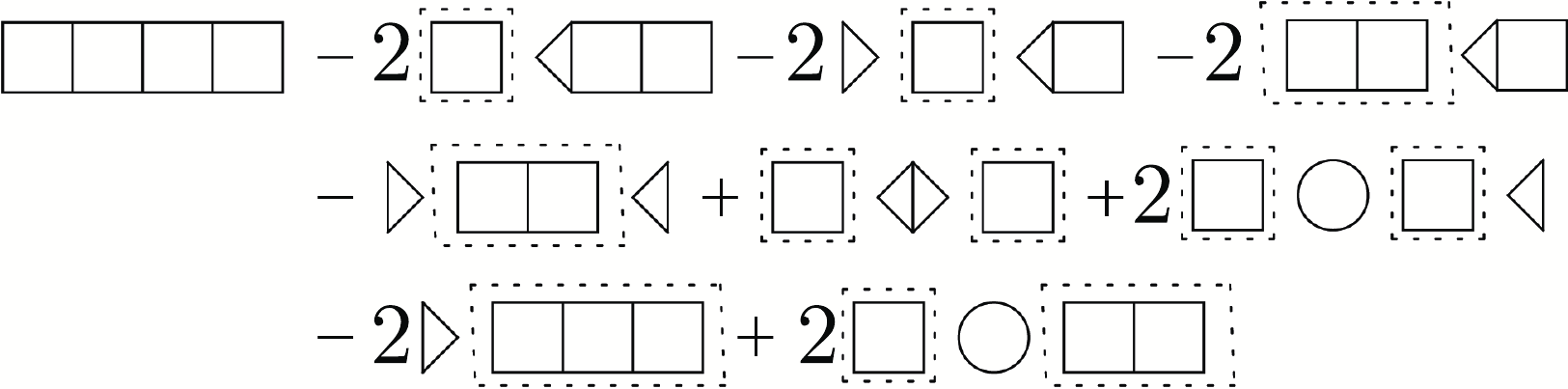}
 \caption{$\R'$-operation for the 4-loop box diagram}\label{4_loop}
 \end{center}
 \end{figure}

The situation repeats itself in the fourth order of PT being even more tricky. In this case, the contribution of the last diagram in Fig.\ref{4_loop}
to the sub-subleading divergence $\sim c_1$ contains the above mentioned $t$ term part. Extracting this part, we get the following 
 sub-subleading divergence $\sim c_1$:
 \beq	
\Delta \Sigma_{sC}^{'trunc}(4-loop) = - \frac{2471 c_1 s^3}{37324800  \epsilon^2}. 
\label{delta_4}
\eeq

At the same time, the subleading term $ \Sigma_{sB}'$ also has the $t$ term contribution.  Excluding this term, we get the truncated expression for $\Sigma_{sB}'$ in 4-loops
\beq 
 \Sigma_{sB}^{'trunc}(4-loop) = - \frac{2471 s^3}{149299200   \epsilon^3}.\label{sB_4}
\eeq
One can see that they coincide and are releated by
\beq 
\Delta \Sigma_{sC}^{'trunc}(4-loop) = c_1z\frac{d\Sigma_{sB}^{'trunc}(4-loop)}{d z}.
\eeq 


These formulae show us the way how the one-loop constant $c_1$ enters  the full answer. It comes from the redefinition of the coupling in a straightforward way
\beq
 z\rightarrow z(1+c_1\epsilon) +  z^2(c_2 -c_1^2/4!)\epsilon^2 + z^3 c_1^3/6! \epsilon^3-z^4 c_1^4/4!6! \epsilon^4+.... \label{power}
\eeq
Note that while expansion of $\Sigma'_A$ starts with the first power of $z$, the extra terms like (\ref{two}) start with $z^2$ and (\ref{three}) with $z^3$, etc. This means that the lowest terms must cancel. This happens when the coefficients of eqs.(\ref{two},\ref{three}) are chosen in a proper way. In fact one can just calculate these coefficients from the requirement of cancelation of the lowest terms. This means that the series (\ref{power}) is actually uniquely fixed.

\section{Conclusion}

Our main concern here was the understanding of the structure of UV divergences in supersymmetric gauge theory with maximal supersymmetry. The example of $D=8$ $N=1$ SYM  theory is instructive and contains all the main features of a class of maximally supersymmetric YM theories. We restricted ourselves to the on-shell scattering amplitudes since after all it is the S-matrix that we want to make finite. 

Our main results can be formulated as follows: 

1) The on-shell scattering amplitudes contain UV divergences that start from one loop and do not cancel (except for the all loop cancellation of bubbles and triangles). 

2) These divergences possess an increasing powers of momenta (derivatives) when increasing the order of PT.  For the four-point scattering amplitude this manifests itself as increasing power of the Mandelstam variables $s$ or $t$.  This means that the theory is not renormalizable by power counting. 

3) Nevertheless, all the higher loop divergences are related to the lower loop ones via explicit pole equations which are the generalization of the RG equations to the case of non-renormalizable theories. The leading divergences are governed by the one-loop counter term, the subleading ones - by the two-loop counter term, etc. This is happening exactly like in the well known case of renormalizable interactions.

4) The summation of the leading and subleading divergences can be performed by solving the generalized RG equations. These solutions obey the characteristic property that they possess an infinite number of poles as functions of  $z=g^2s^2/\epsilon$. This means that they do not have limit when $z\to\infty$ ($\epsilon\to 0$) which would correspond to the finite answer when removing the regularization, i.e. the all loop summation of the leading divergences do not lead to the finite theory.

5) The trouble with non-renormlizable interactions is not that they can not be made finite, but an infinite arbitrariness of the counter terms and, hence, of the finite parts. We have demonstrated  how this arbitrariness may be reduced to the redefinition of the  set of dimensionless couplings $g^2s^2$
(and $g^2t^2$) which are momentum dependent. This is the difference from renormalizable case where one has just one coupling $g^2$. We have not yet find out how to treat these momentum dependent couplings so that to make sense of non-renormalizable theory. 



\section*{Acknowledgements}
This work was supported by the Russian Science Foundation grant \# 16-12-10306.

\end{document}